\documentclass[10pt, conference, letterpaper]{IEEEtran}
\IEEEoverridecommandlockouts
\usepackage{cite}
\usepackage{amsmath,amssymb,amsfonts, amsthm}
\usepackage{algorithm}
\usepackage{algorithmicx}
\usepackage{algpseudocode}
\usepackage{comment}
\usepackage{graphicx}
\usepackage{textcomp}
\usepackage{xcolor}
\usepackage{dsfont}
\usepackage{siunitx}

\usepackage{graphicx}   
\usepackage{multirow}
\usepackage{bigstrut}
\usepackage{array}
\usepackage{subcaption}

\usepackage{lipsum}

\newcommand{\E}[2][]{\mathbb{E}_{#1}\left[#2\right]}

\newtheorem{definition}{Definition}

\newtheorem{theorem}{Theorem}

\newtheorem{proposition}{Proposition}

\def\BibTeX{{\rm B\kern-.05em{\sc i\kern-.025em b}\kern-.08em
    T\kern-.1667em\lower.7ex\hbox{E}\kern-.125emX}}

\begin{document}

\title{Admission Control with Minimal Measurement Cost:
A Best Arm Identification Approach}
\author{\IEEEauthorblockN{Simon Lindståhl\IEEEauthorrefmark{1}\IEEEauthorrefmark{2}, Alexandre Proutiere\IEEEauthorrefmark{1}, Andreas Jonsson\IEEEauthorrefmark{2}}
		\IEEEauthorblockA{\IEEEauthorrefmark{1}KTH Royal Institute of Technology}
		\IEEEauthorblockA{\IEEEauthorrefmark{2}Ericsson Research}
}

\maketitle

\begin{abstract}
	\setlength\parfillskip{0pt plus .75\textwidth}
	\setlength\emergencystretch{1pt}
	In sliced networks, the shared tenancy of slices requires adaptive admission control of data flows, based on measurements of network resources. In this paper, we investigate  the design of measurement-based admission control schemes, deciding whether a new data flow can be admitted and in this case, on which slice. The objective is to devise a joint measurement and decision strategy that returns a {\it correct} decision (e.g., the least loaded slice) with a certain level of confidence while minimizing the measurement cost (the number of measurements made before committing to the decision). We study the design of such strategies for several natural admission criteria specifying what a correct decision is. For each of these criteria, using tools from best arm identification in bandits, we first derive an explicit information-theoretical lower bound on the cost of any algorithm returning the correct decision with fixed confidence. We then devise a joint measurement and decision strategy achieving this theoretical limit. We compare empirically the measurement costs of these strategies, and compare them both to the lower bounds as well as a naive measurement scheme. We find that our algorithm significantly outperforms the naive scheme (by a factor $2-8$). 
	
\end{abstract}

\section{Introduction}\label{sec:intro}

In next generation telecom networks, the network resources will be divided and allocated to multiple slices shared between several slice tenants \cite{9295415}. With limited to no knowledge of the behavior of other tenants, a slice tenant must, in order to uphold certain service guarantees, decide to accept or reject incoming data flows, while adapting to rapidly changing network occupancy levels. This is further complicated by an unclear dependency of the slice occupancy on the resources of individual slice components. The admission control agent must therefore measure the network resources and current utilization before an admission decision can be made, reintroducing a need for \emph{measurement-based admission control} (MBAC), a popular method in the context of call admission control which has recently fallen out of favor \cite{jamin1997measurement, gibbens1997measurement, 779194}. MBAC schemes have the advantage to adapt to uncertainties arising due to the difficulty of characterizing traffic sources or to that of estimating the available resources (in wireless networks, these evolve depending on e.g. user mobility, fading, interference). However, MBAC comes with an inherent cost since a fraction of the resources is used for the measurements. This cost can become substantial as the admission criteria grows in complexity \cite{tahaei2018cost}, especially in system with inherently scarce resources such as wireless systems \cite{camp2008measurement}. 


In this paper, we investigate the design of MBAC strategies in multi-slice networks, where the controller has to decide whether a new data flow can be admitted and if so, on which slice. The controller  has no knowledge about the slice loads, but may gather this knowledge conducting noisy and costly measurements. 
To this aim, it can sequentially measure the traffic handled (over a fixed duration -- a time slot) by a selected slice, and stop whenever it believes it has gathered enough information to come up with a {\it correct} decision with some level of certainty. A correct decision should be to reject the flow if all slices are already fully loaded, or to select one of the slices that has enough available resources if any, given assumptions on the new network flow.  The objective is to devise a joint measurement and decision strategy that returns a correct decision with a certain level of confidence while minimizing the measurement cost (the number of measurements made before committing to the decision). We study the design of such strategies for several natural admission criteria specifying what a correct decision is. These criteria can consist in selecting (i) {\it any} of the slices with available resources, (ii) the most loaded slice with available resources (we refer to this slice as the {\it packing} slice), or (iii) the \emph{least loaded} slice with available resources.

We address the design of joint measurement and decision strategies using the formalism of pure exploration in stochastic Multi-Armed Bandits (MAB). Online exploration algorithms for MAB specify an adaptive sequence of arms (for us, slices) to observe (here, the traffic handled by the selected slice in a slot), a stopping rule indicating when to output a decision, and a decision rule. For each of the aforementioned admission criteria, we first derive an explicit information-theoretical lower bound on the cost of any algorithm returning the correct decision with fixed confidence. We then devise a joint measurement and decision strategy achieving this theoretical limit. We compare empirically the costs of these strategies, and compare these cost both to the lower bound and to a naive sampling strategy. These results allow us to analyze the trade-off between measurement cost and complexity of the proposed admission criteria.

\section{Related work}

{\bf Stochastic bandit} problems have received plenty of attention since they were introduced by Thompson in the 30's and formalized by Robbins in 1952. While bandit problems were initially motivated by clinical trials, they have recently found important applications in the design of protocols and algorithms in communication networks (mostly in cognitive radio systems, see \cite{liu2010, anandkumar2011}
and references therein, or rate adaptation in wireless systems \cite{combes2019}). Most often in bandits, the focus has been on the design of algorithms with low regret \cite{lai1985}. The problem of identifying the best arm using a minimal number of samples is more recent, see \cite{mannor2004, evendar2006} for early work. Algorithms to find the best arm with  minimal sample complexity have been developed in \cite{garivier2016optimal}. Since then, researchers have tried to extend these algorithms to more general {\it pure exploration} problems \cite{degenne2019pure}, such as \cite{garivier2018thresholding} where one searches for the arm with average reward the closest to a given threshold. In this paper, we investigate three novel pure exploration problems, each corresponding to a different admission criterion, and we use the framework developed in \cite{garivier2016optimal} to derive sample complexity lower bounds and to devise optimal algorithms based on these lower bounds. While \cite{degenne2019pure} constructed general lower bounds for such problems, they are often implicit and non-trivial to compute. Furthermore, their algorithm \emph{Sticky Track-and-Stop} cannot typically be implemented without both an explicit form of these bounds and the assumption of Gaussian random variables. By contrast, we provide explicit bounds for our admission criteria as well as an algorithm applicable for a wide class of \linebreak measurement distributions.

{\bf Admission control} methods in the context of network slicing are summarized in \cite{ojijo2020survey}. These methods vary in slice elasticity, inter vs intra-slice admission control, single vs multi-tenant systems, and use both heuristic and optimal methods. None of these methods explicitly take measurement overhead into account. As far as we are aware, this paper proposes the first approach to actually optimize the measurement strategy in admission control. It is worth noting that our admission control problems may seem similar to the problem of dynamic channel assignment in wireless networks, see e.g., \cite{everitt1989performance, zhao2005decentralized,maskery2009decentralized}. However, most existing work in this field concerns the design of Medium Access Control (MAC) protocols (a faster time scale than that of flow arrivals), and most often, channels may take two states only, busy or free.

\section{Models: Dynamics, Admission Control, and Measurement Costs}\label{sec:models}

\begin{figure}[t]
	\centering
	\includegraphics[width=0.45\textwidth]{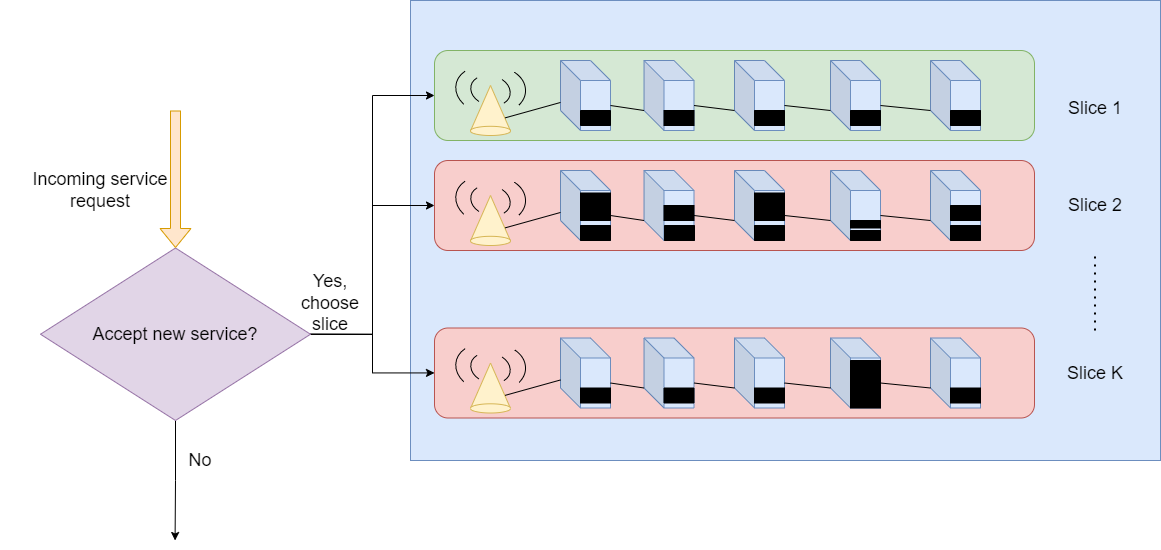}
	\caption{Admission control in sliced networks.}
	\label{fig:sliceadmissioncontrol}
\end{figure}

This section presents our network model, and states our admission control problem. The network consists of $K$ slices of equal capacity.
The network handles traffic {\it flows} or {\it services} generated by end users, and its resources are shared by many users. When a new flow is created, the slice tenant managing these users, or in other words the \emph{controller}, has no knowledge about the current traffic conditions on the various slices, but wishes to select a slice so that the performance guarantees of existing flows in the slice 
remains as high as possible. In this case, this translates to ensuring the loads of all slices remains below some \emph{threshold}. To determine which slice should handle the flow or whether the flow should be rejected, the controller has to measure the traffic intensity on slices. This measurement procedure induces a cost, such as power or bandwidth consumption, that the controller wishes to minimize. We describe this cost minimization problem in detail below, and an outline of the system is found in Figure \ref{fig:sliceadmissioncontrol}. In this figure, the slices are visualized as a chain of virtual network functions (VNFs) depicted as blue boxes, connected to radio over network links with a controller monitoring the network. The shaded areas correspond to the utilization level of the VNFs as consumed by a set of flows. In this case, the correct decision is for the controller to admit into Slice 1 as it is the only available slice.

\subsection{Packet-level dynamics and admission criteria}

Flows are assumed to generate packets according to a stationary process. When for a given slice, the set of accepted flows is fixed, we assume that the aggregate packet arrival process has statistics described, for simplicity, by a single parameter. Time is slotted, and this parameter is defined as the average number of packets arriving in one slot. In this paper, the processes might be Bernoulli (if the slot duration is very small), or Poisson (for the usual Poisson model at packet level in data networks), but extending the results to more sophisticated distributions is simple if so desired. 
For slice $k\in [K]$, we denote by $\mu_k$ as the mean packet arrival rate per slot, fixed during a decision setting, and define $\boldsymbol{\mu}=(\mu_1,\ldots,\mu_K)$. $\mathbb{P}_{\boldsymbol{\mu}}$ (resp. $\mathbb{E}_{\boldsymbol{\mu}}$) denotes the probability distribution (resp. the expectation) of observations when the packet arrival rates are parametrized by $\boldsymbol{\mu}$. 

Assume that a user generates a new flow with known packet arrival rate $r$. Further assume that the current traffic in the network is described by $\boldsymbol{\mu}$. We consider scenarios where accepting the flow should be ideally decided based on $r$ and $\boldsymbol{\mu}$. This happens for example when we wish to guarantee that the average packet delay
of accepted flows remains smaller than a given threshold. For Poisson packet arrivals, the threshold $\gamma$ is obtained simply by plugging $r$, $\boldsymbol{\mu}$, the slice capacity, and packet size statistics in the M/G/1 Pollaczek--Khinchine formula. As a result, the flow should be ideally accepted in slice $k$ only if $\mu_k< \gamma$, in which case, we say that slice $k$ is {\it available}. The flow should be rejected if none of the slices is available. In this paper, when a new flow is created, $\boldsymbol{\mu}$ is unknown and has to be learnt. 

\subsection{Best slice identification problems}

The controller applies a joint measurement and decision strategy to decide whether a newly generated flow can be accepted and if so, on which slice. In each slot $t\ge 1$, we may measure for a selected slice, say $k$, the number of packets $X_k(t)$ handled by the slice in that slot.
For example under the Poisson traffic assumption, the r.v. $X_k(t)$ are i.i.d. with Poisson distribution of unknown mean $\mu_k$. Now a joint measurement and decision strategy consists of three components:
\begin{itemize}
\item[(i)] {\it A sampling strategy.} It specifies, in each slot, the slice to measure. Measurements are taken once per slice and consider end-to-end load, rather than load in individual VNFs (as in Figure \ref{fig:sliceadmissioncontrol}). For $t\ge 1$, denote by $k_t$ and by $X_{k_t}(t)$ the slice probe in slot $t$ and the corresponding number of packets observed. Then $k_t$ depends on past observations, i.e., $k_t$ is ${\cal F}_{t-1}$-measurable where ${\cal F}_t$ is the $\sigma$-algebra generated by $(k_1,X_{k_1}(1),\ldots, k_t,X_{k_t}(t))$.
\item[(ii)] {\it A stopping rule.} It controls the end of the data acquisition phase and is defined as a stopping time $\tau$ with respect to the filtration $(\mathcal{F}_t)_{t\ge 1}$ such that $\mathbb{P}_{\boldsymbol{\mu}}(\tau < \infty) = 1$.
\item[(iii)] {\it A decision rule.} At the end of slot $\tau$, the algorithm returns a decision $\hat{k}(\tau)\in \{0,1,\ldots,K\}$, where $\hat{k}(\tau)=0$ means that the flow is rejected, and $\hat{k}(\tau)=k\ge 1$ is the selected available slice.  $\hat{k}(\tau)$ depends on all the observations made and is hence ${\cal F}_\tau$-measurable.
\end{itemize}

A correct decision is obtained when $\hat{k}(\tau)=0$ if there is no slice with load below the threshold, or when $\hat{k}(\tau)$ is an available slice. There may be multiple available slices, and we can further specify the admission criteria by refining the definition of a correct decision. It is not immediately obvious that any criterion is strictly better than one another, but we will consider algorithms pertaining to each of the following three criteria and compare them to one another. In all scenarios, we denote by ${\cal C}(\boldsymbol{\mu})\subset \{0,1,\ldots,K\}$ the set of correct answers given the server loads $\boldsymbol{\mu}$. We also define $\mu_\star = \min_k \mu_k$ as the smallest load between slices and $k_\star\in\arg\min_k\mu_k$ as the least loaded slice. Finally we let $k^\star \in \arg\max_{k:\mu_k< \gamma}\mu_k$ be the most loaded available slice, defined only when $\mu_\star<\gamma$.
\begin{enumerate}
	
	\item \setlength\parfillskip{0pt plus .75\textwidth}
	\setlength\emergencystretch{1pt} {\it Any available slice.} Under this criterion, we have ${\cal C}(\boldsymbol{\mu})=\{k: \mu_k< \gamma\}$ if $\mu_\star< \gamma$ and \linebreak ${\cal C}(\boldsymbol{\mu})=\{0\}$ otherwise.
\item {\it Packing slice.} Here, we wish to select the most loaded available slice, referred to as the {\it packing} slice. This choice allows us to get a minimum number of active slices, and in some scenarios where the service rates of incoming flows are heterogenous, to reduce the blocking rate. Under this criterion, ${\cal C}(\boldsymbol{\mu})=\{ k^\star \}$ if $\mu_\star< \gamma$ and ${\cal C}(\boldsymbol{\mu})=\{0\}$ otherwise and we denote $\mu^\star = \mu_{k^\star}$.
\item  {\it Least loaded slice.} Selecting the least loaded available slice is also a natural admission criterion, since it will tend to homogenize the loads of the slices, and hence ensure fairness (packet of the various flows experience similar delays) and low packet delay. Here, ${\cal C}(\boldsymbol{\mu})=\{k_\star\}$ if $\mu_\star< \gamma$ and ${\cal C}(\boldsymbol{\mu})=\{0\}$ otherwise. While superficially similar to the problem considered by \cite{garivier2018thresholding}, this criterion differs in the requirement that the slice be available which creates a discontinuity for loads near $\gamma$ and thereby disqualifies the methods considered in that~paper.
\end{enumerate}

Given one of the aforementioned admission criteria, we wish to design algorithms returning a correct answer with a fixed level of certainty. Note that since $\boldsymbol{\mu}$ is unknown and measurements are inherently noisy, it is impossible to surely get a correct answer. We fix $\delta>0$, and target $\delta$-PC ($\delta$-Probably Correct) algorithms, that is, algorithms which are guaranteed to return the correct answer with at least probability $1 - \delta$:

\begin{definition}[$\delta$-PC algorithms] A joint measurement and decision algorithm is $\delta$-PC if and only if for any $\boldsymbol{\mu}$, $\mathbb{P}_{\boldsymbol{\mu}}[\tau<\infty]=1$ and $\mathbb{P}_{\boldsymbol{\mu}}[\hat{k}(\tau)\notin {\cal C}(\boldsymbol{\mu})]\le \delta$.
\end{definition}

The objective is to devise a $\delta$-PC algorithm with minimal expected measurement cost or sample complexity $\mathbb{E}_{\boldsymbol{\mu}}[\tau]$ for the various envisioned admission criteria.

\subsection{Induced flow-level dynamics}

While this paper mainly focuses on devising efficient measurement schemes, it is worth mentioning and studying the impact of the chosen admission criteria on the flow-level performance, i.e., on the flow blocking probabilities. To simplify the discussion below, we assume that the admission decisions are always correct, so that we can focus on the impact of the chosen admission criteria. The deviations caused by the fact that our algorithms may sometimes fail to output a correct decision are assessed numerically in section \ref{sec:num}.   

\smallskip
\noindent
{\it Homogenous flows.} When flows generate packets at the same rate, then all admission criteria lead to the same dynamics at flow-level (the process describing the number of ongoing flows) and hence the same blocking probability (given for example by one of Erlang formulas if flow arrivals are Poisson). In that case, it is best to choose the admission criteria with the minimum measurement cost.

\smallskip
\noindent
{\it Heterogenous flows.} When the flows have different rates, then the selected admission criterion impacts the flow-level dynamics and blocking probabilities. It has been shown that with heterogeneous flows, the steady-state distribution of the population of flows is sensitive to flow size distribution, arrival process and time scale \cite{bonald2003insensitive}, and we cannot analytically characterize the blocking rates. This difficulty arises essentially because with heterogenous flows, the network dynamics are not monotonic in any sense and not reversible \cite{kelly1979reversibility}. As a consequence, it is difficult to predict the behavior of any given admission controller.  We will investigate the trade-off achieved under different admission criteria between blocking probabilities and measurement costs only numerically (see Section \ref{sec:num}).

\section{Best arm identification in admission control}\label{sec:AbbrTheory}

To devise $\delta$-PC algorithms with minimal measurement cost for each admission criterion, we first derive lower bounds on this cost. For a given criterion, we show that the lower bound is the value of an optimization problem, whose solution specifies the optimal measurement process (it characterizes the numbers of times an algorithm with minimal cost should measure each slice before stopping). We then develop algorithms whose sampling and stopping rules perform this optimal measurement process.


\subsection{Lower bounds}

\noindent
{\bf Notations.} To state the lower bounds, we introduce the following notations. Let $\Lambda$ be the $(K-1)$-dimensional simplex $\Lambda=\{ \boldsymbol{w}\in [0,1]^K: \sum_kw_k=1\}$. We denote by $d(a,b)$ the Kullback-Leibler divergence (KL-divergence) between two distributions of the same one-parameter exponential family, parameterized by means $a$ and $b$, respectively. $d_B(a,b)$ denotes this KL-divergence in the case of Bernoulli distributions. In the sequel, to avoid pathological cases where one cannot identify an available slice even with an infinite number of measurements, we assume that $\mu_\star \neq \gamma$. Furthermore, we introduce the information deviation function as $g_{j,k}(x) = d(\mu_k, (\mu_k+x\mu_j)/(1+x)) + x d(\mu_j,  (\mu_k+x\mu_j)/(1+x))$ and its inverse $x_{j,k}(y) = g_{j,k}^{-1}(y)$. 
We use this to introduce the equilibrium function for a set of candidate arms $\mathcal{S}$ and a target arm $k$
$$F_k(y;\mathcal{S}) = \sum_{j \in \mathcal{S}}\frac{d(\mu_k, (\mu_k + x_{j,k}(y)\mu_j)/(1+x_{j,k}(y)))}{d(\mu_j, (\mu_k + x_{j,k}(y)\mu_j)/(1+x_{j,k}(y)))}.$$
%
Let ${\cal  C}^\star(\boldsymbol{\mu})\subseteq {\cal C}(\boldsymbol{\mu})$ be the set of maximizers over $k$ of $\max_{\boldsymbol{w} \in \Lambda} \inf_{\boldsymbol{\lambda}: k \notin {\cal C}(\boldsymbol{\lambda})} \sum_{\ell=1}^K w_\ell d(\mu_{\ell},\lambda_{\ell})$. Following the interpretation in \cite{degenne2019pure}, slices in ${\cal  C}^\star(\boldsymbol{\mu})$ are the {\it easiest} correct answers to identify, and an optimal algorithm should output one of these slices. 

\noindent
{\bf Lower bounds and the optimal measurement process.} Following the approach developed in \cite{garivier2016optimal}, we identify the cost lower bounds, as well as the corresponding optimal fractions of time each slice should be measured. These fractions, denoted by $\boldsymbol{w}^\star\in \Lambda$, depend on whether there is an available slice and on the admission criterion.  The following propositions are established in Appendix \ref{appx:LowerBoundProofs}.   

\begin{proposition} {\bf [No available slice]}
For all three criteria, if $\mathcal{C}(\boldsymbol{\mu})= \{0\}$ (there is no available slice), then any $\delta$-PC algorithm fulfills $\E[\boldsymbol{\mu}]{\tau} \geq T_0(\boldsymbol{\mu})d_B(\delta,1-\delta)$ where
	$$
	T_0(\boldsymbol{\mu}) = \sum_{k = 1}^K d(\mu_k, \gamma)^{-1}.
	$$
	The optimal measurement process is given by, for all $k$, 
		$w^\star_k(\boldsymbol{\mu}) = \frac{d(\mu_k, \gamma)^{-1}}{T_0(\boldsymbol{\mu})}$.
	\label{prop:RejectLowerBound}
\end{proposition}

\begin{proposition} {\bf [Any-available-slice]} For the any-available-slice problem with $\mathcal{C}(\boldsymbol{\mu}) \neq \{0\}$, any $\delta$-PC algorithm fulfills $\E[\boldsymbol{\mu}]{\tau} \geq T_1(\boldsymbol{\mu})d_B(\delta,1-\delta)$
	with
	$
	T_1(\boldsymbol{\mu}) = d(\mu_\star,\gamma)^{-1}.
	$\\
	The optimal measurement process is given by, for all $k$,
		$w_k^\star(\boldsymbol{\mu}) = \mathds{1}_{(k = k^\star)}$.
	\label{prop:AnythingLowerBound}
\end{proposition}

\begin{proposition} {\bf [Packing-slice]} For the packing-slice problem with $\mathcal{C}(\boldsymbol{\mu}) \neq \{0\}$, define $\mathcal{S}_{P} = \{j: \mu_j < \mu^\star\}$ and $z^\star = \min(d(\mu^\star, \gamma), y^\star)$ where $y^\star$ is the unique solution to the equation $F_{k^\star}(y;\mathcal{S}_{P}) = 1$ ($y^\star$ and $z^\star$ are well defined). Any $\delta$-PC algorithm fulfills $\E[\boldsymbol{\mu}]{\tau} \geq T_2(\boldsymbol{\mu})d_B(\delta,1-\delta)$ where 
	\begin{equation}
		T_2(\boldsymbol{\mu}) = \sum_{k: \mu_k > \mu^\star} d(\mu_k, \gamma)^{-1} +
		\frac{1}{z^\star}\sum_{k:\mu_k<\gamma} x_{k,k^\star}(z^\star).
	\end{equation}
	The optimal measurement process is given by, for all $k$,
	\begin{equation}
		w_k^\star(\boldsymbol{\mu}) =\frac{1}{T_2(\boldsymbol{\mu})} \left( \frac{x_{k,k^\star}(z^\star)\mathds{1}_{(\mu_k < \gamma)}}{z^\star} + \frac{\mathds{1}_{(\mu_k > \gamma)}}{d(\mu_k, \gamma)} \right).
		\label{eq:OptimalPackingProportions}
	\end{equation}

	\label{prop:PackingLowerBound}
\end{proposition}

\begin{proposition} {\bf [Least-loaded-slice]} For the least-loaded-slice problem with $\mathcal{C}(\boldsymbol{\mu}) \neq \{0\}$, define $\mathcal{S}_{LL} = \{j: \mu_j > \mu_\star\}$ and $z_\star = \min(d(\mu_\star, \gamma), y_\star)$ where $y_\star$ is the unique solution to the equation $F_{k_\star}(y; \mathcal{S}_{LL}) = 1$ ($y_\star$ and $z_\star$ are well defined). Any $\delta$-PC algorithm fulfills $\E[\boldsymbol{\mu}]{\tau} \geq T_3(\boldsymbol{\mu})d_B(\delta,1-\delta)$ where 
	\begin{equation}
		T_3(\boldsymbol{\mu}) = \frac{1}{z_\star} \sum_{k = 1}^K x_{k, k_\star}(z_\star).
	\end{equation}
	The optimal measurement process is given by, for all $k$,
	\begin{equation}
		w_k^\star(\boldsymbol{\mu}) = \frac{x_{k,k_\star}(z_\star)}{z_\star T_3(\boldsymbol{\mu})}.
		\label{eq:OptimalMinimumProportions}
	\end{equation}

	\label{prop:MinimumLowerBound}
\end{proposition}

\subsection{Track-and-Stop algorithm}

Next, we describe the Track-and-Stop (\textsc{TaS}) algorithm, a generic algorithm that will be instantiated for the three admission criteria, and establish its asymptotic optimality (when $\delta$ approaches 0).

\medskip
\noindent
{\it Sampling rule.} The measurement cost lower bounds and the corresponding optimal measurement process provide the design principle of the sampling rule. We follow the Track-and-Stop framework developed in \cite{garivier2016optimal}: the sampling rule is designed so as to track the optimal fractions $\boldsymbol{w}^\star(\boldsymbol{\mu})$ of time each slice should be measured. Here $\boldsymbol{\mu}$ is unknown, and hence, for the $t$-th measurement, we track $\hat{\boldsymbol{w}}^\star(t):=\boldsymbol{w}^\star(\hat{\boldsymbol{\mu}}(t-1))$ instead, where $\hat{\boldsymbol{\mu}}(t-1)$ are the estimated slice loads from the $(t-1)$-th first measurements. The algorithm will work as long as we can make sure that $\hat{\boldsymbol{\mu}}(t)$ converges to $\boldsymbol{\mu}$ almost surely. To this aim, the sampling rule includes a {\it forced exploration} phase: after $t$ measurements, slices that have not been measured more than $\sqrt{t}$ times are measured. 
If the algorithm is not in a forced exploration phase, it tracks the allocation $\hat{\boldsymbol{w}}^\star(t)$, i.e., it measures the slice $k_t\in \arg\max_k t\hat{w}^\star_k(t)-n_k(t-1)$, where $n_k(t-1)$ is the number of times $k$ has been measured so far. Finally note that the sampling rule depends on the functions $\boldsymbol{\mu}\mapsto \boldsymbol{w}^\star({\boldsymbol{\mu}})$ specified by Propositions \ref{prop:RejectLowerBound} - \ref{prop:MinimumLowerBound} for the various admission criteria.



\medskip
\noindent
{\it Stopping rule.} The stopping rule we use relies on a similar stopping criterion as in all previously devised pure exploration algorithms. Specifically, it is based on the Generalized Likelihood Ratio (GLR) statistics $Z_{k,k'}(t)$ evaluating the probabilities that given the observations, the targeted correct answer is $k=k_E(t)$ or $k'$, see details in \cite{garivier2016optimal} and \cite{degenne2019pure}. We stop when these GLR are large enough. The resulting statistical test can summarized by comparing $Q(t):=\inf_{\boldsymbol{\lambda}: \hat{k}(t) \notin \mathcal{C}(\boldsymbol{\lambda})} \sum_{k=1}^K n_{k}(t-1) d(\hat{\mu}_{k}(t-1), \lambda_k)$ to an exploration threshold $f_\delta(t)$ appropriately chosen.  

\medskip
\noindent
{\it Decision rule.} When the algorithm stops measuring after $\tau$ measurements, it returns the slice $k_E(\tau)$. The pseudo-code of the algorithm is presented in Algorithm \ref{alg:TaS}. There, $\mathbf{n}(t)$ denotes the vector counting the number of times each slice has been measured up to time $t$ (such that $\sum_{k=1}^K n_{k}(t) = t$).

\begin{algorithm}[t]
	\caption{Track-and-Stop (\textsc{TaS})}
	\label{alg:TaS}
	\begin{algorithmic}
		
		\State {\bf Input:} Oracle functions $\boldsymbol{\mu}\mapsto \mathbf{w}^\star(\boldsymbol{\mu})$ and $\boldsymbol{\mu}\mapsto {\cal C}^\star(\boldsymbol{\mu})$
		\State {\bf Initialization:} $\hat{\boldsymbol{\mu}}(0) = 0$, $\mathbf{n}(0) = 0$
		\For{$t=1,...$}
		\State ${k}_E(t)\leftarrow  \min {\cal C}^\star(\hat{\boldsymbol{\mu}}(t-1))$
		\State $Q(t)\leftarrow \inf_{\boldsymbol{\lambda}: {k}_E(t) \notin \mathcal{C}(\boldsymbol{\lambda})} \sum_{k=1}^K n_{k}(t-1) d(\hat{\mu}_{k}(t-1), \lambda_k)$
		\If{$Q(t) > f_\delta(t)$}
		\State Stop and return ${k}_E(t)$
		\EndIf
		\State $\hat{\mathbf{w}}(t)\leftarrow  \mathbf{w}^\star(\hat{\boldsymbol{\mu}}(t-1))$
		\If{$\exists k: n_{k}(t-1) < \sqrt{t}$}
		\State $k_t \leftarrow \arg\min_k n_{k}(t-1)$
		\Else
		\State $k_t \leftarrow \arg \max_k t \hat{w}_{k}(t) - n_{k}(t-1)$
		\EndIf
		\State Measure slice $k_t$ and observe $X_{k_t}(t)$
		\State  $\mathbf{n}(t) \leftarrow \mathbf{n}(t-1) + {\bf e}_{k_t}$
		\State $\hat{\boldsymbol{\mu}}(t)\leftarrow \hat{\boldsymbol{\mu}}(t-1) + \frac{1}{n_{k_t}(t)}(X_{k_t}(t) - \hat{\mu}_{k_t}(t-1)) {\bf e}_{k_t}$
		\EndFor
	\end{algorithmic}
\end{algorithm}


\medskip
The following theorem, proved leveraging results from \cite{degenne2019pure} in Appendix \ref{appx:UpperBoundProofs}, establishes the asymptotic (as $\delta$ goes to 0) optimality of \textsc{TaS} for the any-available-slice problem, up to a factor 2. 

\begin{theorem}
	Let \textsc{TaS} be instantiated for any of our admission criteria with input functions $\boldsymbol{\mu}\mapsto \mathbf{w}^\star({\boldsymbol{\mu}})$  given by (\ref{eq:OptimalAnythingProportions}), (\ref{eq:OptimalPackingProportions}) and (\ref{eq:OptimalMinimumProportions}), respectively. Select the exploration threshold equal to $f_\delta(t) =\log(C t^2/ \delta)$ with $C$ such that $C \geq e \sum_{t=1}^\infty (\frac{e}{K})^K\frac{(\log(Ct^2)\log(t))^K}{t^2}$. Then, \textsc{TaS} is $\delta$-PC, and its sample complexity satisfies on criteria $i$: for any $\boldsymbol{\mu}$ such that $k_\star$ is unique,
	\begin{equation}
		\limsup_{\delta \to 0} \frac{\mathbb{E}_{\boldsymbol{\mu}}[\tau]}{d_B(\delta,1-\delta)} \leq 2T_i(\boldsymbol{\mu}).
	\end{equation}
	\label{the:AlgResults}
\end{theorem}

We conclude this section by remarking that for $\boldsymbol{\mu}$ such that $k_\star$ is not unique, we would need to add a component to the algorithm to avoid the oscillation of $k_E(t)$ between these slices with minimal load. This is done in \cite{degenne2019pure} by introducing a {\it sticky} component to the algorithm. We could follow this idea, but for simplicity and clarity of the paper, prefer to restrict our attention to the cases where $k_\star$ is unique. 
\section{Simulations}\label{sec:num}

We have run simulations to illustrate the performance of our joint measurement and decision strategies under the various admission criteria. To do so, we fixed the system load and invested the measurement costs of our algorithm using the different admission criteria. Then, we investigated to what extent the constraints are violated and compared the sampling efficiency with a naive sampling algorithm. Finally, we studied the dynamical behavior of the system given that different admission criteria gives rise to different admission control behavior.

\subsection{Packet level dynamics and local optimality}
We studied a system with $K=8$ slices of equal slice resources. The flows we are interested in are services in the range of Conversational Voice, Conversational Video and Live Gaming. Conversational Voice has a packet frequency of about $50$ packets/s \cite{cisco2016voip}. Conversational video sends a bitrate of about 1500 kbps \cite{ibm20xxbitrate} and has a packet size of about 1500 bytes \cite{sengupta2017movidiff} which translates to a packet frequency of about 125 packets per second. Finally, Live Gaming has a typical packet interarrival time of $40$ ms which translates to $25$ packets/s \cite{che2012packet}. In this first case of study, we therefore assumed that each slice could hold 24 flows,
that each UE sent, on average, $50$ packets per second and that in each measurement we measured the number of packets sent during a slot of length $20$ ms \footnote{$20$ ms is chosen arbitrarily to have each UE send about $1$ packet per slot, but we can see similar results with other slot lengths.}. Thus, the admission threshold $\gamma$ was fixed equal to $\gamma = 24 \times 50 \times 0.02=24$ packets/slot, corresponding to a bit rate of 14 Mbps.

The traffic intensity $\mu$ on each slice was generated randomly from a truncated uniform multinomial distribution according to the following parameters:

\begin{enumerate}
	\item The total load is fixed according to the scenarios
	\begin{itemize}
		\item \emph{Low load} with an average of $17$ packets/slot/slice,
		\item \emph{Medium load} with an average of $23$ packets/slot/slice and
		\item \emph{High load} with an average of $30$ packets/slot/slice.
	\end{itemize}  
	\item The probability of any unit load being assigned to a slice is uniform over slices.
\end{enumerate}

\begin{table*}[t]
	\centering
	\begin{tabular}{r|ccc|ccc|ccc}
		& \multicolumn{3}{c|}{{\bf Any-available-slice}}  & \multicolumn{3}{c|}{{\bf Packing-slice}} & \multicolumn{3}{c}{{\bf Least-loaded-slice}} \\
		Scenario 
		&
		\textsc{TaS} & Uniform & Lower bound & \textsc{TaS} & Uniform & Lower bound & \textsc{TaS} & Uniform & Lower bound \\
		\hline
		Low load	 & 12.2 & 27.7 & 0.843 & 808 & 2800 & 245 & 390 & 1020 & 139 \\
		Medium load & 28.4 & 105 & 2.88  & 1790 & 5480 & 571 & 565 & 1350 & 180\\
		High load & 571 & 3522 & 92.6 & 897 & 3640 & 261 & 897 & 2310 & 290\\
		
	\end{tabular}
	\caption{Measurement costs in slots (averaged over 100 runs -- confidence intervals are not shown due to space constraints but are typically small, with radius of the order of $5\%$ of mean).}
	\label{tab:PacketResults}
\end{table*}

Note that since the intensity is generated by a multinomial distribution, it only takes integer values. In the first and second scenarios, an available slice is always available by the pigeon hole principle, and so, we expected a greater discrepancy between the different admission criteria in measurement cost. In the third scenario, an available slice is not always available and so we expected the difference to be smaller.
We implemented \textsc{TaS} in Python 3.7 under the three admission criteria, and compared their performance to that of a naive algorithm using the same stopping and decision rules but with a sampling rule picking slices in a round-robin manner. This benchmark is useful as it allows us to see the impact of only our intelligent sampling rule, removing the impact of confidence levels and other performance guarantees. The level of confidence for the stopping rule was fixed to $\delta = 0.01$. Each algorithm was tested on 100 independent runs, and in each run the traffic intensity $\boldsymbol{\mu}$ was regenerated. The results are shown in Table \ref{tab:PacketResults}, including the averaged lower bounds from Section \ref{sec:AbbrTheory} for comparison.

\begin{figure}[t]
	\centering
	\includegraphics[width=0.4\textwidth]{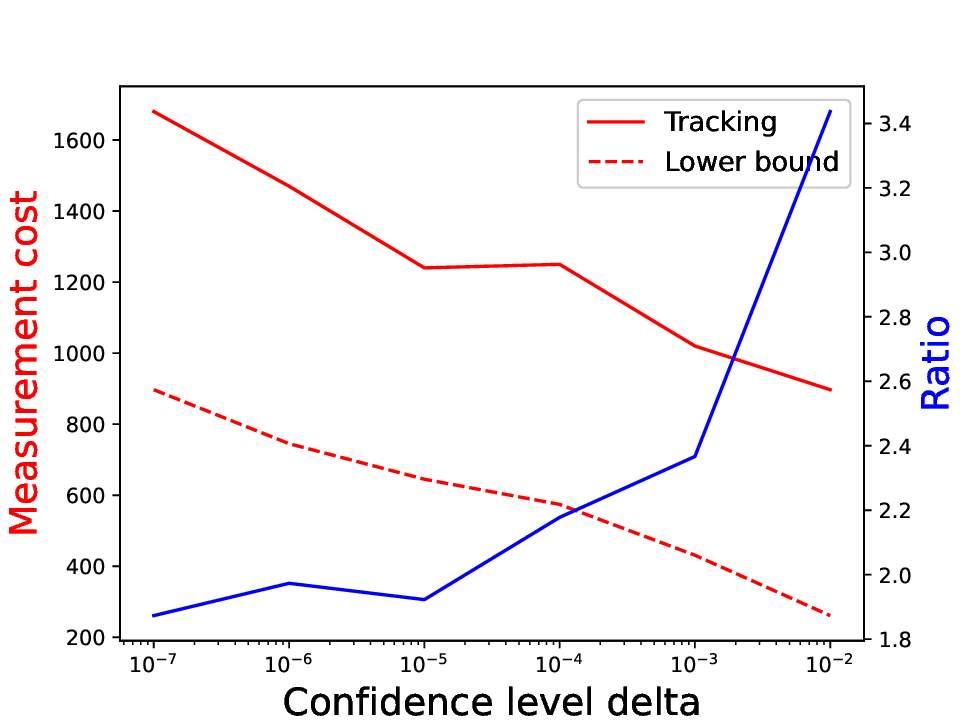}
	\caption{The measurement costs as a function of the confidence parameter $\delta$. The red lines depict the measurement cost and lower bound while the blue line represents the ratio between the two.}
	\label{fig:confidence_ratio}
\end{figure}

Observe that the problems with different admission criteria have different difficulties. As expected, the any-available-slice criterion leads to a much lower measurement cost, except for the high load scenario where there is often no or a single available slice. In all scenarios, \textsc{TaS} significantly outperforms a naive algorithm using uniform sampling: the improvements in the measurement cost are by a factor $2$ to $8$ on all problems. \textsc{TaS} measurement costs are not so close to the lower bound, an effect we attribute to the moderate confidence level \linebreak of $\delta = 0.01$. To verify this, we evaluated the measurement cost, as compared to the lower bound, for a variety of values of $\delta$ in the range $[10^{-7}, 10^{-2}]$ with the packing-slice criterion and high load. These costs and lower bounds, along with the ratio between the two are shown in Figure \ref{fig:confidence_ratio}. We see that the measurement cost increases sub-linearly with $\log(1/\delta)$ and that it converges to approximately $1.8$ when $\delta \to 0$, which is consistent with Theorem \ref{the:AlgResults}.

\subsection{Measurement costs and blocking rates averaged over flow dynamics}
\begin{figure*}[t]
	\centering
	\begin{subfigure}{0.32\textwidth}
		\includegraphics[width=\textwidth]{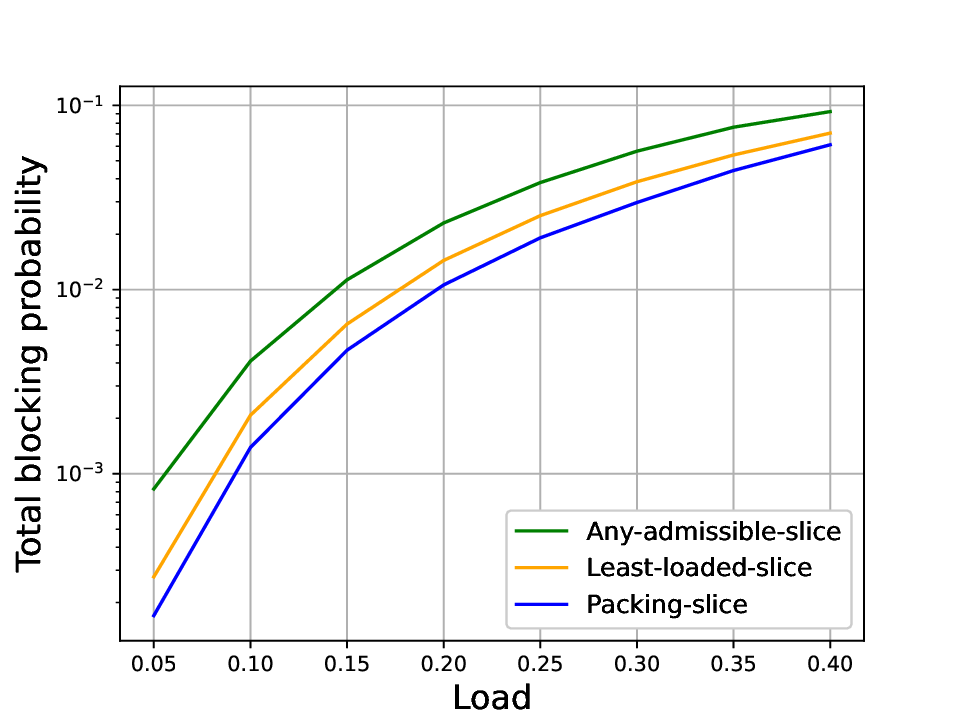}
		\caption{Overall blocking probability}
	\end{subfigure}
	\begin{subfigure}{0.32\textwidth}
		\includegraphics[width=\textwidth]{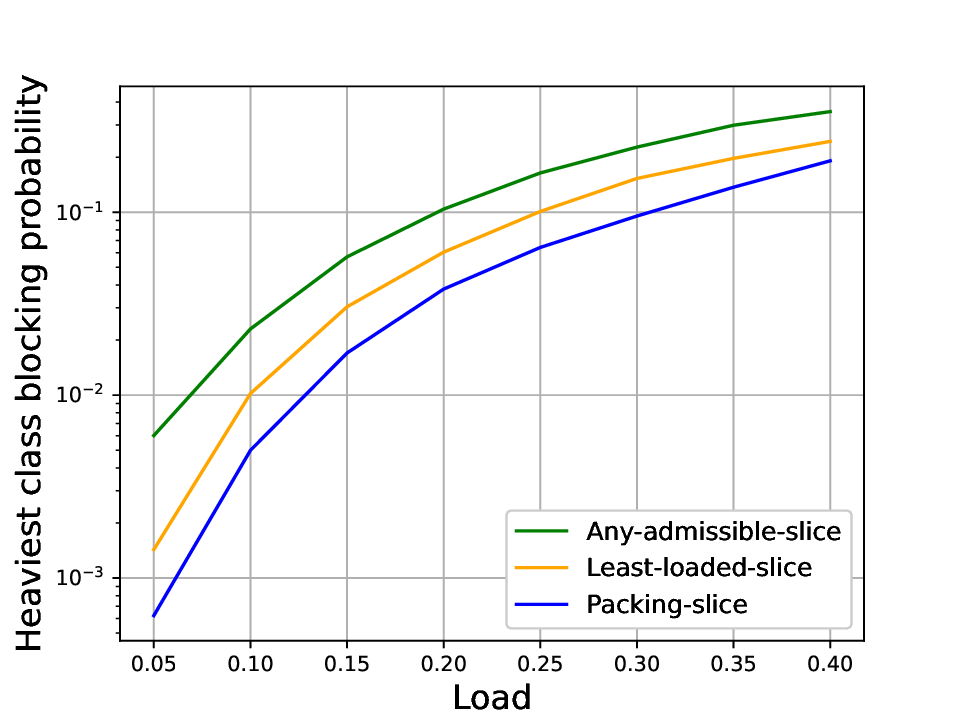}
		\caption{Blocking probability of high-rate flows}
	\end{subfigure}
	\begin{subfigure}{0.32\textwidth}
		\includegraphics[width=\textwidth]{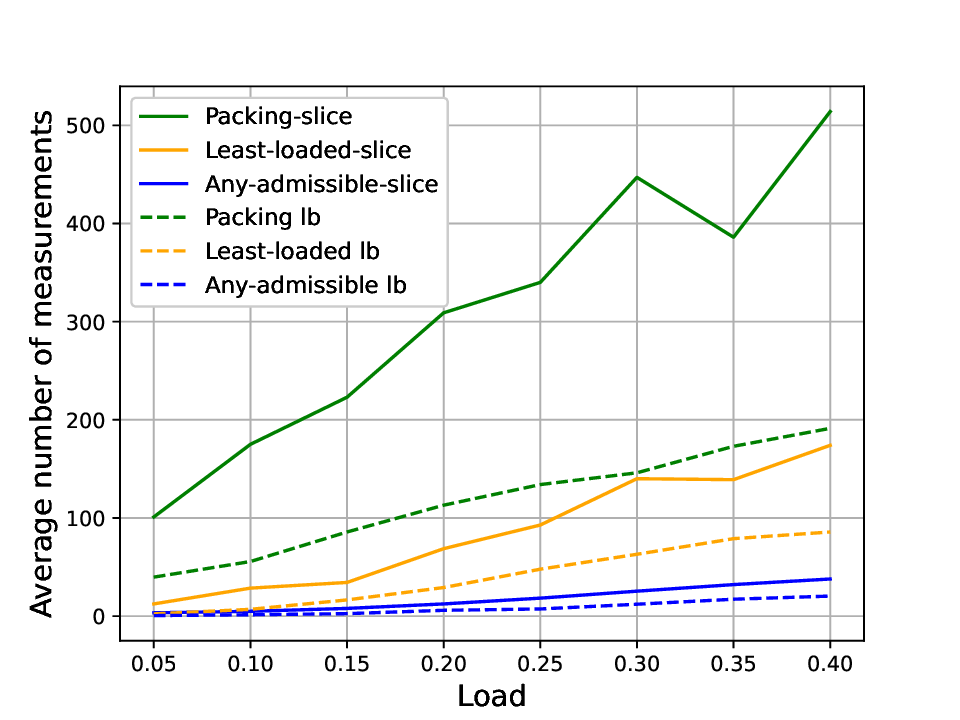}
		\caption{Number of measurements}
	\end{subfigure}
	\caption{Flow-level performance under the various admission criteria and \textsc{TaS}. (a) Blocking probability averaged over all flows vs. load; (b) Blocking probability of flows with the highest rate (10) vs. load; (c) Measurement cost vs. load (solid lines report the performance of \textsc{TaS}, and dashed lines our averaged lower bounds).}
\label{fig:AlgorithmComparison}
\end{figure*}

Next, we account for our simulations on flow-level dynamics. Flows were generated according to a Poisson process of intensity $\bar{\lambda}$ (flows per time unit). When a flow arrives, its packet arrival rate $r$ was chosen uniformly at random in $\{1, ..., 10\}$ per slot. We used $K=3$ slices, each of capacity 15.5 packets per slot. A flow of rate $r$ can be accepted on a slice $k$ only if $\mu_k < \gamma = 15.5 -r$ (the admission threshold depends on the rate $r$). Flow durations were exponentially distributed with mean of 1 time unit. Note that the time unit is assumed to be much larger than the duration of a slot, so that the population of flows can be assumed to be fixed over a slot when measurements were conducted. Overall, the {\it load} of the system used was $\rho = {\bar{\lambda}\bar{r}\over 15.5 K}$ where $\bar{r}=5.5$ is the average flow rate. For \textsc{TaS}, the confidence level $\delta$ was set to $0.01$. We examined all admission criteria: any-available-slice, packing-slice and least-loaded-slice. For the measurement cost, we include the lower bounds $T_i(\mu)$  from Section \ref{sec:AbbrTheory}, averaged over the flow states $\mu$ seen by the controller across the experiment.  The results can be found in Figure \ref{fig:AlgorithmComparison}.

Figure \ref{fig:AlgorithmComparison} highlights the trade-off between blocking probabilities and measurement costs. The any-available-slice criterion offers the lowest measurement cost but this comes at the expense of a higher blocking probability. The overall blocking probability is 2 to 5 times greater under the any-available-slice criterion than under the packing-slice criterion. The latter offers the lowest blocking rates but has a high measurement cost -- almost 10 times greater compared to the any-available-slice criterion. It seems that the least-loaded-slice criterion yields the best trade-off between blocking rates and measurement costs.

Another important result, not presented in these figures, is that during our experiments, there was no occasion during which the system was overloaded. This suggests that \textsc{TaS} is more conservative compared to what the targeted confidence level $\delta = 0.01$ imposes.

\section{Conclusions}\label{sec:conclusion}

In this paper, we investigated the problem of admission control in network slicing where before admitting or rejecting a flow, the slice utilization of several slices needs to be measured. Inspired by Best Arm Identification methods, we designed a framework to allow for robust admission control with confidence guarantees. We applied this framework to devise optimal joint measurement and admission schemes realizing three different admission criteria. We verified, using simulations, the optimality of our algorithms as compared to the lower bounds and showed their advantage over naive measurement methods. In this paper, we assumed that the unknown parameters, driving the admission decisions and implicitly learnt by our algorithms, just dictate the loads of the slices. In practice, in real sliced networks, there might be other types of uncertainty (e.g. unknown flow rate or unknown slice capacity), and we plan to extend our methods and results to deal with these additional uncertainties.

%
%


\begin{appendices}
\section{Lower bound proofs}\label{appx:LowerBoundProofs}

\subsection{Proof of Propositions \ref{prop:RejectLowerBound} and \ref{prop:AnythingLowerBound}}
Recall that $\Lambda$ is the simplex of dimension $(K - 1)$. For any correct answer $\ell \in \mathcal{C}(\boldsymbol{\mu})$, denote by $\text{Alt}(\ell) := \{ \boldsymbol{\lambda}: \ell \notin \mathcal{C}(\boldsymbol{\lambda}) \}$, and denote by $D(\boldsymbol{w}, \boldsymbol{\mu}, \boldsymbol{\lambda}) := \sum_{k=1}^Kw_kd(\mu_k,\lambda_k)$.
By Theorem 1 in \cite{degenne2019pure}, any $\delta$-PC algorithm must fulfill $\E[\boldsymbol{\mu}]{\tau} \geq T(\boldsymbol{\mu})d_B(\delta, 1 - \delta)$ where
$$
T(\boldsymbol{\mu})^{-1} = \max_{{w} \in \Lambda} \max_{\ell \in \mathcal{C}(\boldsymbol{\mu})}\inf_{\boldsymbol{\lambda}: \ell \in Alt(\boldsymbol{\lambda})} \sum_{k=1}^K w_k d(\mu_k, \lambda_k).
$$
 We therefore wish to solve the max-min problem
$$
\boldsymbol{w}^\star(\boldsymbol{\mu}) \in \arg\max_{\boldsymbol{w} \in \Lambda} \max_{\ell \in \mathcal{C}(\boldsymbol{\mu})} \inf_{\boldsymbol{\lambda} \in \text{Alt}(\ell)} D(\boldsymbol{w}, \boldsymbol{\mu}, \boldsymbol{\lambda}).
$$
%

We begin with the case $\mu_\star > \gamma$. Then $\mathcal{C}(\boldsymbol{\mu}) = \{0\}$ by definition of our admission criteria. We have that $\text{Alt}(0) = \{\boldsymbol{\lambda}:\exists k \in [K]: \lambda_k < \gamma\}$, and from this set, we can restrict ourselves to studying only $\boldsymbol{\lambda}^{(k)}$ such that $\lambda_{m}^{(k)}(\nu) = \nu \mathds{1}_{(m=k)} + \mu_m\mathds{1}_{(m \neq k)}$ with $\nu < \gamma$ (only a single arm $k$ is changed compared to $\boldsymbol{\mu}$). Indeed, for any instance $\boldsymbol{\lambda} \in \text{Alt}(\boldsymbol{\mu})$ and any weights $\boldsymbol{w} \in \Lambda$, there exists $k$ and $\nu$ such that $D(\boldsymbol{w}, \boldsymbol{\mu}, \boldsymbol{\lambda}) \geq D(\boldsymbol{w}, \boldsymbol{\mu}, \boldsymbol{\lambda}^{(k)}(\nu))$. 
It follows that for any $\boldsymbol{w} \in \Lambda$,  $\inf_{\boldsymbol{\lambda} \in \text{Alt}(\ell)}D(\boldsymbol{w}, \boldsymbol{\mu}, \boldsymbol{\lambda}) = \min_{k\in[K]} D(\boldsymbol{w},\boldsymbol{\mu},\boldsymbol{\lambda}^{(k)}(\gamma)) = \min_{k\in[K]}w_kd(\mu_k, \gamma)$. We note then that for maximizing $\boldsymbol{w}=\boldsymbol{w}^\star(\boldsymbol{\mu})$, it must be true that $D(\boldsymbol{w}^\star(\boldsymbol{\mu}),\boldsymbol{\mu},\boldsymbol{\lambda}^{(k)}(\gamma)) =D(\boldsymbol{w}^\star(\boldsymbol{\mu}),\boldsymbol{\mu},\boldsymbol{\lambda}^{(m)}(\gamma))\ \forall k \in [K],m \in [K]$. Adding the condition $\sum_{k=1}^Kw_k = 1$, this yields a linear equation system which is easily solved with
\begin{equation}
	w^\star_k(\boldsymbol{\mu}, 0) = \frac{d(\mu_k, \gamma)^{-1}}{\sum_{j=1}^K d(\mu_j, \gamma)^{-1}}
\end{equation}
and therefore, by the above lower bound, $T_0(\boldsymbol{\mu}) = \sum_{k=1}^K d(\mu_k, \gamma)^{-1}$, which concludes this case. However, this case also extends easily to the Packing-slice and Least-Loaded-slice problem, since it is simple to show that the set of confusing problems $\text{Alt}(0)$ remains the same for these.

Next, we study the any-available-slice case with $\mu_\star < \gamma$. Take any arm $\ell \in \mathcal{C}(\boldsymbol{\mu})$,  with $\mu_\ell < \gamma$. For this case, $\text{Alt}(\ell)$ are all problems $\boldsymbol{\lambda}$ such that $\lambda_\ell > \gamma$. Similarly as before, it is easy to see that $ \inf_{\boldsymbol{\lambda} \in \text{Alt}(\ell)} D(\boldsymbol{w}, \boldsymbol{\mu}, \boldsymbol{\lambda}) =D({\bf w}, \boldsymbol{\mu}, \boldsymbol{\lambda}^{(\ell)}(\gamma)) = w_\ell d(\mu_\ell,\gamma)$. Furthermore, this expression is maximized under $\boldsymbol{w} \in \Lambda$ and $\ell \in \mathcal{C}(\boldsymbol{\mu})$ by $\boldsymbol{w}^\star(\boldsymbol{\mu})$ with $w^\star_k(\boldsymbol{\mu}) = \mathds{1} (k = \ell)\ \forall k \in [K]$ and $\ell = k_\star(\boldsymbol{\mu})$. Therefore, we see that Proposition \ref{prop:AnythingLowerBound} holds with $T_1(\boldsymbol{\mu}) = d(\mu_\star, \gamma)^{-1}$ as above. This concludes this case and the proof.

\subsection{Proof of Proposition \ref{prop:PackingLowerBound}}
		We will assume that $\mu_\star <\mu^\star < \gamma$ (for the case $\mu_\star = \mu^\star < \gamma$, refer to the proof of Propositions \ref{prop:RejectLowerBound} and \ref{prop:AnythingLowerBound}). In this case, $\mathcal{C}(\boldsymbol{\mu}) = \{k^\star(\boldsymbol{\mu})\}$, with $k^\star(\boldsymbol{\mu})$ unique by hypothesis. We notice that $\text{Alt}(k^\star(\boldsymbol{\mu}))$ can be written as the union of three sets, $\text{Alt}(k^\star(\boldsymbol{\mu})) \subseteq A_1 \cup A_2 \cup A_3$ where

	\begin{enumerate}
		\item $A_1 := \{\boldsymbol{\lambda}: \exists k: \mu_k > \gamma,\ \lambda_k < \gamma\}$,
		\item $A_2 := \{\boldsymbol{\lambda}: \exists k: \mu_k < \mu^\star, \lambda_{k^\star(\boldsymbol{\mu})} < \lambda_k < \gamma \}$,
		\item $A_3 := \{\boldsymbol{\lambda}: \lambda_{k^\star(\boldsymbol{\mu})} > \gamma \}$.
	\end{enumerate}
Hence,	%
	$$
	\inf_{\boldsymbol{\lambda} \in \text{Alt}(k^\star(\boldsymbol{\mu}))}D(\boldsymbol{w}, \boldsymbol{\mu}, \boldsymbol{\lambda}) = \inf_{\boldsymbol{\lambda} \in A_1 \cup A_2 \cup A_3}D(\boldsymbol{w}, \boldsymbol{\mu}, \boldsymbol{\lambda}).
	$$
	We will first find the value of $\inf_{\boldsymbol{\lambda} \in A_i}D(\boldsymbol{w}, \boldsymbol{\mu}, \boldsymbol{\lambda})$ for $i \in \{1,2,3\}$. This value is
	
	\begin{enumerate}
		\item for $A_1$, $\min_{k: \mu_k > \gamma} w_k d(\mu_k, \gamma)$,
		\item for $A_2$, $\min_{k: \mu_k < \mu^\star}: w_{k^\star(\boldsymbol{\mu})}I_{w_{k^\star(\boldsymbol{\mu})}/(w_k + w_{k^\star(\boldsymbol{\mu})})}(\mu^\star, \mu_k)$, 
		\item  and for $A_3$, $w_{k^\star(\boldsymbol{\mu})}d(\mu^\star, \gamma)$.
	\end{enumerate}
	
	Let $D^\star(\boldsymbol{w}, \boldsymbol{\mu}, \text{Alt}(k^\star(\boldsymbol{\mu})))$ be the minimum of these three expressions. Since the sets of modified arms are non-overlapping between $A_1$ and $A_2 \cup A_3$, we see that for each arm $k$ with $\mu_k > \gamma$ the expression $w_kd(\mu_k, \gamma)$ must be equal and identical to $D^\star(\boldsymbol{w}, \boldsymbol{\mu}, \text{Alt}(k^\star(\boldsymbol{\mu})))$ for any maximizing $\boldsymbol{w}$, and thereby, for these arms $w^\star_k(\boldsymbol{\mu}, k^\star(\boldsymbol{\mu})) = D^\star(\boldsymbol{w}^\star(\boldsymbol{\mu}, k^\star(\boldsymbol{\mu})), \boldsymbol{\mu}, \text{Alt}(k^\star(\boldsymbol{\mu})))d(\mu_a, \gamma)^{-1}$.
	
	 For $A_2$, the best proportions $w^\star_k$ are known from Best Arm Identification problems \cite{garivier2016optimal} and can be found as $w^\star_k = \frac{x_{k,k^\star(\boldsymbol{\mu})}(y^\star) D^\star(\boldsymbol{w}, \boldsymbol{\mu}, \text{Alt}(k^\star(\boldsymbol{\mu})))}{y^\star}$ for all $k$ with $\mu_k < \gamma$, where $x_{k,k^\star(\boldsymbol{\mu})}(y)$ and $y^\star$ are defined in the proposition.
	
	Now, if $y^\star \leq d(\mu^\star, \gamma)$ then $\frac{D^\star(\boldsymbol{w}, \boldsymbol{\mu}, \text{Alt}(k^\star(\boldsymbol{\mu}))) d(\mu^\star, \gamma)}{y^\star} \geq D^\star(\boldsymbol{w}, \boldsymbol{\mu}, \text{Alt}(k^\star(\boldsymbol{\mu})))$ and so these proportions maximize the expression $\min_{\boldsymbol{\lambda} \in A_2  \cup A_3} D(\boldsymbol{w}, \boldsymbol{\mu}, \boldsymbol{\lambda})$. 
	
	If instead $y^\star > d(\mu^\star, \gamma)$, by convexity of $D^\star(\cdot, \boldsymbol{\mu}, \text{Alt}(k^\star(\boldsymbol{\mu})))$, we see that $w_{k^\star(\boldsymbol{\mu})}=\frac{D^\star(\boldsymbol{w}, \boldsymbol{\mu}, \text{Alt}(k^\star(\boldsymbol{\mu})))}{d(\mu_1, \gamma)}$ is maximizing and so, since the expression $I_{w_{k^\star(\boldsymbol{\mu})}/(w_a + w_{k^\star(\boldsymbol{\mu})})}(\mu^\star, \mu_a)$ must be identical to $\frac{D^\star(\boldsymbol{w}, \boldsymbol{\mu}, \text{Alt}(k^\star(\boldsymbol{\mu})))}{w_{k^\star(\boldsymbol{\mu})}}$ for all arms $k$ with $\mu_k < \gamma$, we obtain $w^\star_k = \frac{D^\star(\boldsymbol{w}, \boldsymbol{\mu}, \text{Alt}(k^\star(\boldsymbol{\mu})))}{d(\mu_1, \gamma) x_{k,k^\star(\boldsymbol{\mu})}(d(\mu^\star, \gamma))}{d(\mu^\star, \gamma)}$.
	In summary, recalling that $z^\star = \min(d(\mu^\star, \gamma), y^\star)$, we find that $w^\star_k(\boldsymbol{\mu}, k^\star(\boldsymbol{\mu})) = \frac{D^\star(\boldsymbol{w}^\star(\boldsymbol{\mu}, k^\star(\boldsymbol{\mu})), \boldsymbol{\mu}, \text{Alt}(k^\star(\boldsymbol{\mu}))) x_{k,k^\star(\boldsymbol{\mu})}(z^\star)}{z^\star}\ \forall k:\ \mu_k < \gamma$.
	
	Finally, we make use of the fact that $\sum_{k} w_ak= 1$, and find that 
	\begin{align*}
		D^\star(\boldsymbol{w}^\star&(\boldsymbol{\mu}), \boldsymbol{\mu}, \text{Alt}(k^\star(\boldsymbol{\mu}))) =\\
		&= \left(\sum_{k: \mu_k > \gamma}d(\mu_k, \gamma)^{-1} + \frac{1}{z^\star} \sum_{k:\mu_k< \gamma} x_{k,k^\star(\boldsymbol{\mu})}(z^\star) \right)^{-1}.
	\end{align*}
	But since these proportions $\boldsymbol{w}^\star(\boldsymbol{\mu}, k^\star(\boldsymbol{\mu}))$ are maximizing, we obtain by Theorem 1 in \cite{degenne2019pure} that any $\delta$-PC algorithm must be sampled in expectation at least $T_2(\boldsymbol{\mu})d_B(\delta,1-\delta)$ times, with $T_2(\boldsymbol{\mu}) = D^\star(\boldsymbol{w}^\star(\boldsymbol{\mu}, k^\star(\boldsymbol{\mu})), \boldsymbol{\mu}, \text{Alt}(k^\star(\boldsymbol{\mu})))^{-1} = \sum_{k: \mu_k > \gamma } d(\mu_k, \gamma)^{-1} + \frac{1}{z^\star}  \sum_{k:\mu_k< \gamma} x_{k,k^star}(z^\star)$. This concludes the proof.
	
\subsection{Proof of Proposition \ref{prop:MinimumLowerBound}}
We recall that $\mathcal{C}(\boldsymbol{\mu}) = \{k_\star(\boldsymbol{\mu})\}$ with $k_\star(\boldsymbol{\mu})$ unique by hypothesis. As in the proof of Proposition \ref{prop:PackingLowerBound}, we note that $\text{Alt}(k_\star(\boldsymbol{\mu})) \subseteq A_2 \cup A_3$ with $A_2 = \{\boldsymbol{\lambda}: \exists k: \lambda_k < \lambda_{k_\star(\boldsymbol{\mu})}\}$ and $A_3 = \{\boldsymbol{\lambda}: \lambda_{k_\star(\boldsymbol{\mu})} > \gamma\}$, and furthermore
$$
\inf_{\boldsymbol{\lambda} \in \text{Alt}(k^\star(\boldsymbol{\mu}))}D(\boldsymbol{w}, \boldsymbol{\mu}, \boldsymbol{\lambda}) = \inf_{\boldsymbol{\lambda} \in  A_2 \cup A_3}D(\boldsymbol{w}, \boldsymbol{\mu}, \boldsymbol{\lambda}).
$$
Next, like before we find that $\inf_{\boldsymbol{\lambda} \in A_2}D(\boldsymbol{w}, \boldsymbol{\mu}, \boldsymbol{\lambda}) = \min_{k: \mu_k>\mu_\star}w_{k_\star(\boldsymbol{\mu})} I_{w_{k_\star(\boldsymbol{\mu})}/(w_k + w_{k_\star(\boldsymbol{\mu})})}(\mu_\star, \mu_k)$ and\\ $\inf_{\boldsymbol{\lambda} \in A_3}D(\boldsymbol{w}, \boldsymbol{\mu}, \boldsymbol{\lambda}) = w_{k_\star(\boldsymbol{\mu})} d(\mu_\star, \gamma)^{-1}$.

As in the proof of Proposition \ref{prop:PackingLowerBound}, if $y_\star \leq d(\mu_\star, \gamma)$, then the optimal proportions are those of a Best Arm Identification problem and $\boldsymbol{w}^\star_k(\boldsymbol{\mu}, k_\star(\boldsymbol{\mu})) = \frac{x_{k,k_\star(\boldsymbol{\mu})}(y_\star)}{y_\star T_3(\boldsymbol{\mu})}$ with $T_3(\boldsymbol{\mu}) := \frac{1}{y_\star}\sum_{k=1}^K x_{k,k_\star(\boldsymbol{\mu})}(y_\star)$ for all $k$ with $y_\star$ and $x_{k,k_\star(\boldsymbol{\mu})}$ defined as in Section \ref{sec:AbbrTheory}.

Otherwise, by convexity of $D(\cdot, \boldsymbol{\mu}, \text{Alt}(k_\star(\boldsymbol{\mu}))))$ and by $w_{k_\star(\boldsymbol{\mu})}d(\mu_\star, \gamma) = D(\boldsymbol{w}, \boldsymbol{\mu}, \text{Alt}(k_\star(\boldsymbol{\mu})))$, we find that $w^\star_{k_\star(\boldsymbol{\mu})}(\boldsymbol{\mu}) = d(\mu_\star, \gamma)^{-1} D(\boldsymbol{w}, \boldsymbol{\mu}, \text{Alt}(k_\star(\boldsymbol{\mu})))$. Since then $d(\mu_\star, \gamma)^{-1} D(\boldsymbol{w}, \boldsymbol{\mu}, \text{Alt}(k_\star(\boldsymbol{\mu})))g(w_k/w_{k_\star(\boldsymbol{\mu})}) = D(\boldsymbol{w}, \boldsymbol{\mu}, \text{Alt}(k_\star(\boldsymbol{\mu})))$ it follows that $w^\star_k(\boldsymbol{\mu}) = w^\star_{k^\star(\boldsymbol{\mu})}(\boldsymbol{\mu})x_{\star,k}(d(\mu_\star, \gamma))$ for all $k$. Utilizing $\sum_{k=1}^Kw_k = 1$ and defining in this case $T_3(\boldsymbol{\mu}) :=D(\boldsymbol{w}, \boldsymbol{\mu}, \text{Alt}(k_\star(\boldsymbol{\mu})))^{-1}= \sum_{k=1}^K d(\mu_\star, \gamma)^{-1}x_{\star,k}(d(\mu_\star, \gamma))$, Equation \eqref{eq:OptimalMinimumProportions} follows. The remainder of Proposition \ref{prop:MinimumLowerBound} follows from Theorem 1 in \cite{degenne2019pure}.

\section{Upper bound proof and correctness of \textsc{TaS}}\label{appx:UpperBoundProofs}

In this section we prove Theorem \ref{the:AlgResults}. The proof logic closely follows that of \cite{garivier2016optimal}. We will show that this result is more generally applicable. 

First, take $\epsilon > 0$ and define the event $\mathcal{E}_T(\epsilon) = \bigcap_{t=\sqrt{T}}^T \{||\hat{\boldsymbol{\mu}}(t) - \boldsymbol{\mu}||_\infty < \xi(\epsilon)\}$, where $\xi(\epsilon)> 0$ is the greatest value such that 
\begin{align*}
||\boldsymbol{\mu}' - \boldsymbol{\mu}||_\infty < \xi(\epsilon) \implies &\mathcal{C}^\star(\boldsymbol{\mu}) \subseteq \mathcal{C}(\boldsymbol{\mu}'),\\
& ||\boldsymbol{w}^\star(\boldsymbol{\mu}') - \boldsymbol{w}^\star(\boldsymbol{\mu})||_\infty < \epsilon.
\end{align*}
Such $\xi(\epsilon)$ necessarily exists by continuity of $\boldsymbol{w}^\star$ (Proposition 6 of \cite{garivier2016optimal}, Theorem 4 of \cite{degenne2019pure}). Given a sufficiently small value of $\epsilon$, (say $\epsilon < \epsilon'(\boldsymbol{\mu})$ for some function $\epsilon'$) it holds that on $\mathcal{E}_T(\epsilon)$, $\hat{k}(t) \in \mathcal{C}^\star(\boldsymbol{\mu})$. We define $k_E(\boldsymbol{\mu}) = k_\star(\boldsymbol{\mu})$ in the case of packing-slice and $k_E(\boldsymbol{\mu}) = k^\star(\boldsymbol{\mu})$ in the case of any-available-slice or least-loaded-slice as the target of the true problem. On $\mathcal{E}_T(\epsilon)$ for $t \geq \sqrt{T}$, 
$$Q(t) = t \inf_{\boldsymbol{\lambda} \in \text{Alt}(k_E(\boldsymbol{\mu}))} D(\boldsymbol{n}(t)/t, \hat{\boldsymbol{\mu}}(t), \boldsymbol{\lambda}) .$$

Further, by Lemma 20 of \cite{garivier2016optimal}, for \textsc{TaS} there exists $T_\epsilon$ independent of $\delta$ such that for $T\geq T_\epsilon$ on $\mathcal{E}_T(\epsilon)$ it holds that $\forall t \geq \sqrt{T}$, $||\boldsymbol{n}(t)/t-\boldsymbol{w}^\star(\boldsymbol{\mu})||_\infty \leq 3(K-1)\epsilon$. Introducing $C^\star_\epsilon(\boldsymbol{\mu}) = \inf_{\boldsymbol{\mu}': ||\boldsymbol{\mu}' - \boldsymbol{\mu}||_\infty \leq \xi(\epsilon), \boldsymbol{w}': ||\boldsymbol{w}' - \boldsymbol{w}^\star(\boldsymbol{\mu})||_\infty \leq 3(K-1)\epsilon} D(\boldsymbol{w}', \boldsymbol{\mu}', \text{Alt}(\boldsymbol{\lambda}))$ it then follows that on $\mathcal{E}_T(\epsilon)$ for every $T \geq T_\epsilon$ and $t \geq \sqrt{T}$
$$
Q(t) \geq t C^\star_\epsilon(\boldsymbol{\mu}).
$$
Then, on $\mathcal{E}_T(\epsilon)$ and with $T \geq T_\epsilon$, we have
\begin{align*}
	\min(\tau_\delta, T) &\leq \sqrt{T} + \sum_{t=1}^T\mathds{1}_{(\tau_\delta > t)}
	\leq \sqrt{T} + \sum_{t=1}^T\mathds{1}_{(Q(t) \leq f_\delta(t))}\\
	&\leq \sqrt{T} + \sum_{t=1}^T\mathds{1}_{(tC^\star_\epsilon(\boldsymbol{\mu}) \leq f_\delta(T))}\leq \sqrt{T} +\frac{f_\delta(T)}{C^\star_\epsilon(\boldsymbol{\mu})} 
\end{align*}
Introducing $\tau_0(\delta) = \inf\{T: T\geq \sqrt{T} + \frac{f_\delta(T)}{C^\star_\epsilon(\boldsymbol{\mu})}\}$, it follows that for every $T\geq \max(\tau_0(\delta), T_\epsilon)$, $\mathcal{E}_T(\epsilon) \subseteq (\tau_\delta \leq T)$. As such, we have
\begin{align*}
	\E[\boldsymbol{\mu}]{\tau_\delta}&= \sum_{T=1}^\infty \mathbb{P}(\tau_\delta > T)\leq \tau_0(\delta) + T_\epsilon + \sum_{T=1}^\infty \mathbb{P}(\mathcal{E}_T^c(\epsilon))\\
	& \leq \tau_0(\delta) + T_\epsilon + \sum_{T=1}^\infty BT\exp(-CT^{1/8})
\end{align*}
where the final inequality follows from Lemma 19 in \cite{garivier2016optimal} for some constants $B$ and $C$ depending on $\epsilon$ and $\boldsymbol{\mu}$ but not $\delta$. Following the remainder of the proof step by step, we find that 
\begin{equation*}
	\limsup_{\delta \to 0} \frac{\E[\boldsymbol{\mu}]{\tau_\delta}}{\log(1/\delta)} \leq \limsup_{\delta \to 0} \frac{\tau_0(\delta)}{\log(1/\delta)} \leq \frac{2(1+\eta)}{C_\epsilon^\star(\boldsymbol{\mu})}
\end{equation*} 
for some $\eta > 0$ of our choice. By Theorem 4 of \cite{degenne2019pure}, the function $(\boldsymbol{w}, \boldsymbol{\mu})\mapsto D(\boldsymbol{w}, \boldsymbol{\mu}, \text{Alt}(\ell))$ is continuous on $\Lambda \times \mathcal{M}$ for any $\ell \in \mathcal{C}(\boldsymbol{\mu})$, so letting  $\epsilon \to 0$, $C^\star_\epsilon(\boldsymbol{\mu}) \to D(\boldsymbol{w}^\star(\boldsymbol{\mu}),\boldsymbol{\mu}, \text{Alt}(k_E(\boldsymbol{\mu})))$. But by Propositions \ref{prop:AnythingLowerBound}, \ref{prop:PackingLowerBound} and \ref{prop:MinimumLowerBound}, we see that this value is identical to $T_1(\boldsymbol{\mu})^{-1}$, $T_2(\boldsymbol{\mu})^{-1}$ and $T_3(\boldsymbol{\mu})^{-1}$ respectively, and we obtain the desired bound by letting $\eta \to 0$.

For the $\delta$-PC property, we will use a proof strategy similar to that of Proposition 12 in \cite{garivier2016optimal}. We will show that whenever an error occurs,  $f_\delta(t) < Q(t) < \sum_{k=1}^K n_k(t)(\hat{\mu}_k(t), \mu_k)$. Then, $\delta$-correctness follows from Theorem 2 in \cite{magureanu2014lipschitz}.

The first inequality is an immediate consequence of the statement of \textsc{TaS} (because errors occur only when the agent stops), so we focus on the second. For any-available-slice, packing-slice or least-loaded-slice, all errors can be divided into two categories, and we will show the inequality for both: 

(i) For some arm $\ell,\ \hat{\mu}_\ell(t) < \gamma < \mu_\ell$ or $\mu_\ell < \gamma < \hat{\mu}_\ell(t)$. In either case, for this to be considered an error it must be true that $Q(t) \leq n_\ell(t) d(\hat{\mu}_\ell(t), \gamma) < n_\ell(t) d(\hat{\mu}_\ell(t),\mu_\ell) \leq \sum_{k=1}^K n_k(t)d(\hat{\mu}_k(t),\mu_k)$, which concludes this category.

(ii) For some arms $j, \ell:$ $\hat{\mu}_j(t) < \hat{\mu}_\ell(t)$ and $\mu_\ell < \mu_j$. Defining $\alpha = n_j(t)/(n_j(t) + n_\ell(t))$, if this is considered an error, it can be shown (se the proof of Proposition 12 in \cite{garivier2016optimal}) that,  $Q(t) < n_j(t) d(\hat{\mu}_j(t), \alpha \hat{\mu}_j(t) + (1-\alpha)\hat{\mu}_\ell(t)) + n_\ell(t) d(\hat{\mu}_\ell(t), \alpha \hat{\mu}_j(t) + (1-\alpha)\hat{\mu}_\ell(t)) \leq n_j(t)d(\hat{\mu}_j(t),\mu_j) + n_\ell(t)d(\hat{\mu}_\ell(t),\mu_\ell) \leq \sum_{k=1}^K n_k(t)d(\hat{\mu}_k(t),\mu_k)$. This concludes the proof.

\end{appendices}

\bibliographystyle{IEEEtran}  
\bibliography{references}

\end{document}